\documentclass[a4paper,10pt]{article}
\usepackage[margin=1in]{geometry}
\usepackage{authblk}
\usepackage{graphicx}
\usepackage{color}
\usepackage{hyperref}
\usepackage{pdflscape}
\newcommand\blfootnote[1]{%
  \begingroup
  \renewcommand\thefootnote{}\footnote{#1}%
  \addtocounter{footnote}{-1}%
  \endgroup
}

\begin{document}

\title{
Corruption Risk in Contracting Markets:\\ A Network Science Perspective}


\author[1]{Johannes Wachs}
\author[2]{Mih\'aly Fazekas}
\author[3]{J\'anos Kert\'esz}

\affil[1]{Chair for Computational Social Sciences and Humanities, RWTH Aachen}
\affil[2]{School of Public Policy, Central European University}
\affil[3]{Department of Network and Data Science, Central European University}

\maketitle

\begin{abstract}
We use methods from network science to analyze corruption risk in a large administrative dataset of over 4 million public procurement contracts from European Union member states covering the years 2008-2016. By mapping procurement markets as bipartite networks of issuers and winners of contracts we can visualize and describe the distribution of corruption risk. We study the structure of these networks in each member state, identify their cores and find that highly centralized markets tend to have higher corruption risk. In all EU countries we analyze, corruption risk is significantly clustered. However, these risks are sometimes more prevalent in the core and sometimes in the periphery of the market, depending on the country. This suggests that the same level of corruption risk may have entirely different distributions. Our framework is both diagnostic and prescriptive: it roots out where corruption is likely to be prevalent in different markets and suggests that different anti-corruption policies are needed in different countries. 

\end{abstract}

\section{Introduction}
Ever\blfootnote{Address correspondence to: johannes.wachs@cssh.rwth-aachen.de. This article is based on a chapter of the doctoral dissertation of Johannes Wachs.} since states have existed, they have sought to count and track their citizens in order to tax and control them~\cite{scott1998seeing}. Modern states record their own activities in great detail. Just as the digital traces of individuals contain valuable insights about everything from their health to their socioeconomic status~\cite{pappalardo2016analytical}, the administrative Big Data maintained by the state can be used to analyze its successes and failures~\cite{kim2014big}. One such failure that is surprisingly persistent even among economically advanced and democratic states is corruption, which we frame as  the ``deliberate restriction of open and fair access to public resources for the benefit of connected actors''~\cite{mungiu2015quest}. Corruption is indeed a significant failure of governance, as it has been shown to slow growth~\cite{mauro1995corruption} and innovation~\cite{rodriguez2014quality} and to subvert democracy~\cite{stockemer2013bribes},  while compounding inequality~\cite{gupta2002does}.

Despite its importance corruption is difficult to measure because individuals engaging in corruption obviously want to keep it a secret. Ground truth examples of convicted corrupt actors are hard to come by and anyway form a biased sample: large-scale corruption often goes unpunished because the state, including law enforcement and the judiciary, have been captured by corrupt interests~\cite{mungiu2006corruption}. For these and other reasons corruption is traditionally measured via surveys which suffer from well-documented shortcomings~\cite{hawken2009you,olken2009corruption}. 

The significant amount of administrative data collected by governments presents an opportunity to study corruption risk from a new perspective using new tools of data science. In this paper we mine a big administrative dataset with over 4 million public procurement contracts in the EU for insights on the organization of corruption. Public procurement, the process by which governments purchase goods, services and construction works from the private sector, accounts for up to 20\% of GDP~\cite{oecdprocurement} and is known to be vulnerable to corruption~\cite{fazekas2016corruption}. We proxy for corruption risk at the contract level with a binary indicator that is 1 if there was no competition for the contract. Such single-bidder contracts have been shown to predict significant risk of corruption~\cite{klavsnja2015corruption,charron2017careers}. Aggregated to the country level, the rate of single bidding correlates significantly with commonly used survey-based measures of corruption or predict overpricing on the auction level. We mention here that there are a variety of possible indicators for corruption risk based on the details of the processes of procurement contracts, however, the indicator based on single bidding is a the best choice for a one-parameter description of corruption~\cite{fazekas2016corruption}.

With micro-level data in hand, we apply the tools of network science to map and analyze the distribution and structure of corruption risks in different European countries. We represent markets as bipartite networks of the issuers (public institutions, sometimes referred to as buyers) and winners (firms, sometimes referred to as suppliers) of procurement contracts. These bipartite networks have many qualitative characteristics in common with other empirical networks~\cite{newman2003structure} that distinguish them from random networks. For instance they have heterogeneous degree distributions and significant local correlations.

These networks are more than just maps of markets. We argue that they provide an ideal framework for studying the organization of corruption. There are many reasons to consider corruption as a fundamentally networked phenomenon. The most significant disclosed examples of corruption, for instance the recent scandal involving the Brazilian state-owned oil company Petrobras~\cite{watts2017operation,ribeiro2018dynamical}, involve hundreds of individuals, firms, and institutions. The Petrobras scheme entailed the exchange of billions of dollars in bribes and kickbacks for the award of public contracts. The size of the conspiracy, of which nearly 100 conspirators - including the former president of Brazil - have been convicted, suggests that it involved a sophisticated and collective effort. Likewise, network studies of organized crime~\cite{calderoni2011strategic} and terrorism~\cite{krebs2002mapping} demonstrate that illegal activity tends to leave complex traces that networks are especially suited to describe and explain. 

We describe the distribution of corruption risk in different procurement markets from two perspectives: the extent to which it is centralized, that is how prevalent it is between actors in the core of the market versus its periphery; and how clustered it is, describing the extent to which corruption risk is bunched in different parts of the network. We find that in some EU countries corruption risk is more common in the core of the market, while in others it is more common in its periphery. In every country in our database corruption risk is clustered, though the magnitude of this effect varies significantly. Our results are compared to realistic null models, which take into account the overall magnitude of corruption risk as well the propensity of governments in different countries to purchase different goods and services. Our findings have significant implications for both the academic study of corruption and practical anti-corruption efforts. Observing that corruption seems to be organized in different ways suggests that different anti-corruption strategies may be effective in different contexts.

The paper is organized as follows. Section 2 surveys related work, including recent work on the measurement of corruption risk via public procurement contracts. Section 3 describes the data and corruption risk indicator. Section 4 introduces the network measures of the distribution of corruption risk in markets. Section 5 concludes and suggests ideas for future research.

\section{Related Work}
Big data has been applied to a variety of social and economic problems in domains including health~\cite{murdoch2013inevitable}, scientific research~\cite{sinatra2016quantifying}, urban mobility~\cite{pappalardo2015using,szell2018crowdsourced}, and development~\cite{hilbert2016big}. The digitization of our social and professional lives have facilitated most of this work, providing researchers access to high resolution trace data about human behavior and activities. Often the data analyzed is ``found data'' - data which was not originally collected for the purpose of the research. Data collected from public administrative databases, for example public procurement, falls into this category~\cite{connelly2016role}. 

The study of corruption using big administrative data is a new and evolving area of research with its own challenges. It represents a major departure from previous empirical studies of corruption which leverage perception-based surveys for macro studies and experiments at the micro level. Below we briefly survey these three broad branches of research on corruption.

The most prominent and longest running survey-based measures of corruption are Transparency International's ~\textit{Corruption Perceptions Index} (CPI)~\cite{TI_CPI} and the World Bank's ~\textit{Worldwide Governance Indicators} (WGI)~\cite{WB_WGI}, both available since the 1990s. Both measures, quantifying corruption and quality of government at the national level, are composite indicators, mixing general population and expert surveys about perceptions and experience. The measures are highly correlated ($\rho > $ .9) and are designed to be consistent over time. The significant complexity of weighing or including components and correlating observations year to year has lead some researchers to criticize the validity and broad application of these measures~\cite{hawken2009you,heywood2014close}. More recent attempts to resolve some of these issues or to create more consistent measures include the \textit{Bayesian Corruption Index}~\cite{standaert2015divining}, the ~\textit{Varieties of Democracy} (V-DEM) indicator of political corruption~\cite{coppedge2018v}, and the Quality of Government Institute's (QoG) ~\textit{European Quality of Government Index} (EQI)~\cite{charron2014regional}. The latter index is calculated at the regional level, enabling the study of corruption perceptions at the sub-national level.

Experimental studies of corruption have the advantage of allowing researchers to directly measure causes or levels of corruption in specific contexts by precise specification. For example, to study the influence of culture on corruption Cameron et al.~\cite{Cameron2009} observed individuals from different cultures playing games with both an economic incentive to cheat and a mechanism to punish cheaters. Interestingly, they found greater variation in the propensity to punish corruption than to engage in it across cultures. Weisel and Shalvi~\cite{Weisel2015} observe pairs of individuals playing a dice-rolling game in which pairs can increase their payout by cheating as a team, and find that pairs tend to cheat more often than individuals playing a similar game alone.

These studies have the clear limitation that they happen in artificial environments. Some experiments are carried out in the field, though coming with significant higher costs and risks. Perhaps the most famous of these relating to corruption are Olken's field experiments in Indonesia~\cite{olken2007monitoring,olken2009corruption}. Olken designed a series of interventions to test the effects of both audits and grassroots organization on corruption outcomes in 600 Indonesian villages during a national road construction project. Crucially, Olken had the resources to independently assess the expected cost of the roads in order to estimate actual observed corruption in each village. He found that pre-announced audits reduce corruption, but that community organizing induced no change. Such studies are a gold-standard in causal inference, but suffer from significant costs and limitations in scope.

In recent years researchers have increasingly turned to big data to quantify and study corruption in the public sector. These studies tend to leverage found data collected and curated by governments for other purposes. They rely heavily on new trends for open government or e-government~\cite{kornberger2017bureaucracy}. As citizens increasingly demand transparency and public-sector use of information and communication technologies proliferates, it is reasonable to expect that more data will become available in the future~\cite{bertot2010using}. Data in the United States on lobbying~\cite{borisov2015corporate} and campaign contributions~\cite{bonica2014mapping,traag2016complex} have been used to measure the influence of money in politics. Internationally, large scale firm ownership data reveals how firms avoid taxes~\cite{garcia2017uncovering}. Crowdsourced data on convicted public officials, for instance extracted from Wikipedia, has been used to study the emergence of systemic corruption~\cite{ribeiro2018dynamical}.

As mentioned in the introduction, our study leverages big data on public procurement. The scale and scope of procurement and its status as a key interface between the public and private sectors have made it a popular topic of research. How do researchers quantify corruption using public procurement data? Fieldwork suggests that corrupt officials steer contracts to favored firms by restricting competition, insuring high profits~\cite{fazekas2016objective}. In general, traces of the strategies used to restrict competition can be extracted from administrative data on the contracting process, enabling researchers to score individual contracts for corruption risk. The most general indicator leverages the outcome of the competition directly: whether the contract attracted a single bidder. Single-bid contracts represent a clear corruption risk.

Single-bid contracting rates have been used as an effective proxy for corruption in various contexts, including studies on the relationship between corruption and political incumbency~\cite{klavsnja2015corruption}, the importance of meritocracy in bureaucratic outcomes~\cite{charron2017careers}, the impact of social networks on local corruption~\cite{bergh2019municipally,wachs2019social}, and the effect of campaign contributions on corruption~\cite{fazekas2018institutional}. Procurement-based corruption risk indicators have the advantage that they apply to micro-level transactions, enabling the study of corruption at multiple scales. Several previous studies study the distribution of corruption risk in procurement markets represented as networks. One such study finds that repeated interactions are significantly related to corruption risk~\cite{popa2019uncovering}, while another demonstrates how procurement markets change when corruption becomes endemic~\cite{fazekas2016corruption}.

Despite the significant interest in corruption as a problem and the proliferation of public procurement-based big data indicators, we are not aware of research that explores the distributional and structural properties of corruption in different countries based on the patterns of interactions between public and private actors. This gap in the literature is surprising given that social scientists have been categorizing countries by the organization of corruption in their public sectors for decades~\cite{klitgaard1988controlling,johnston2005syndromes}. We fill this gap by leveraging the tools of network science, providing an analytic framework to both study the distribution of corruption risks in countries and to provide tailor suggestions on how to combat it.

\section{Data and Framework}

\subsection{Procurement Data}
Our analytic framework measures corruption risk at a transaction
level using data from public procurement contracts. We collected data on all public procurement contracts published in Tenders Electronic Daily\footnote{\url{https://ted.europa.eu}} (TED), the official journal of public procurement contracts of the European Union, from 2008 to 2016. Both calls for tenders and the announcement of their award are published in TED. EU law requires that any public procurement contract exceeding an estimated value of 5.2 million Euro for works and 135 thousand Euro for services and supplies must be published on TED. TED estimates that over 400 billion Euro of total contract value is published in its pages every year, accounting for nearly 3\% of EU GDP. Though the high thresholds exclude a significant number of contracts from our data, using only data from TED maximizes the international comparability of our analyses.

Our final dataset consists of 4,098,771 contracts awarded in 2008-2016. We consider 26 member states of the EU, excluding Luxembourg because of the relatively small number of contracts awarded there and Croatia because it only joined the EU in 2013. We also exclude contracts awarded by EU institutions (the Commission, Parliament, Council) as we aim to compare countries. 

We processed the dataset to deduplicate the identities of each contract's issuing buyer (i.e. public institution such as a ministry or city hall) and supplying winner (i.e. a private firm). Given that our analytic approach requires an accurate map of the interactions between issuers and winners, we developed a pipeline to maximize the accuracy of our deduplication, following the approach of Christen~\cite{christen2012data}. For each country, we preprocessed the text data for each entity, used machine learning to select both optimal string similarity measures and blocking methods, and selected a clustering threshold maximizing accuracy on a manually labeled subsample using the Dedupe computer software~\cite{gregg2015dedupe}. To give an example, the number of unique suppliers of contracts in France decreased from 364,125 to 200,584. For a more detailed summary of the procedure see \cite{wachsthesis}. 

Besides the issuing buyer and winning supplier of each contract, we were able to extract several additional pieces of information, including the year the contract was awarded, the Common Procurement Vocabulary (CPV) code - a EU-wide taxonomy of procurement contracting goods and services~\cite{cpvreport}, and the number of bidders competing for the contract. We use the later information to score each contract for corruption risk.

\subsection{Corruption Risk}

We quantify the corruption risk of procurement contracts using a binary indicator tracking whether the contract attracted only a single bid in its competition. The use of single bidding as a signal for corruption risk or ``undetected fraud'' has recently been suggested by the European Court of Auditors~\cite{auditors2019fraud}. Nevertheless, it is certainly not the case that any single instance of single-bidding can be used as evidence for corrupt behavior. For instance, it may be the case that the government is purchasing a niche good or service with few suppliers on the market or under emergency circumstances - certainly it is more likely for such contracts to be awarded to a single bidder.

Two aspects of our data mitigate these limitations to the validity of single-bidding as a signal of corruption risk. The first is that the high threshold of contract values insure both visibility and interest from the private sector. Secondly, the null models we create to benchmark our measures consider the CPV code of the contract, distinguishing between contracts for different categories of goods and services, such as medicine and furniture.

\begin{figure}[t]
  \includegraphics[width=\textwidth]{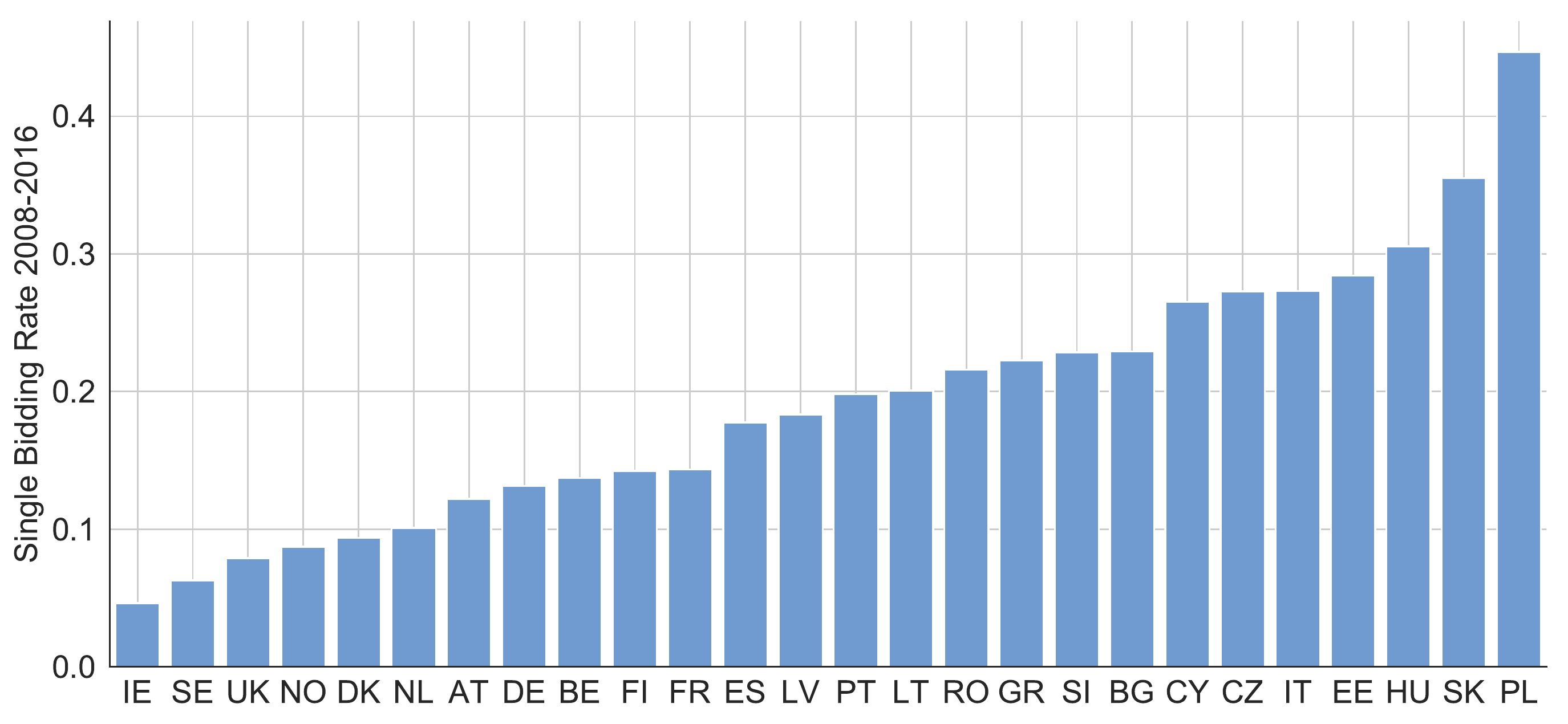}
\caption{Single-bidding rates on procurement contracts on TED, 2008-2016 by country.}
\label{fig:country_sb_rates}
\end{figure}

\begin{figure}[t]
  \includegraphics[width=\textwidth]{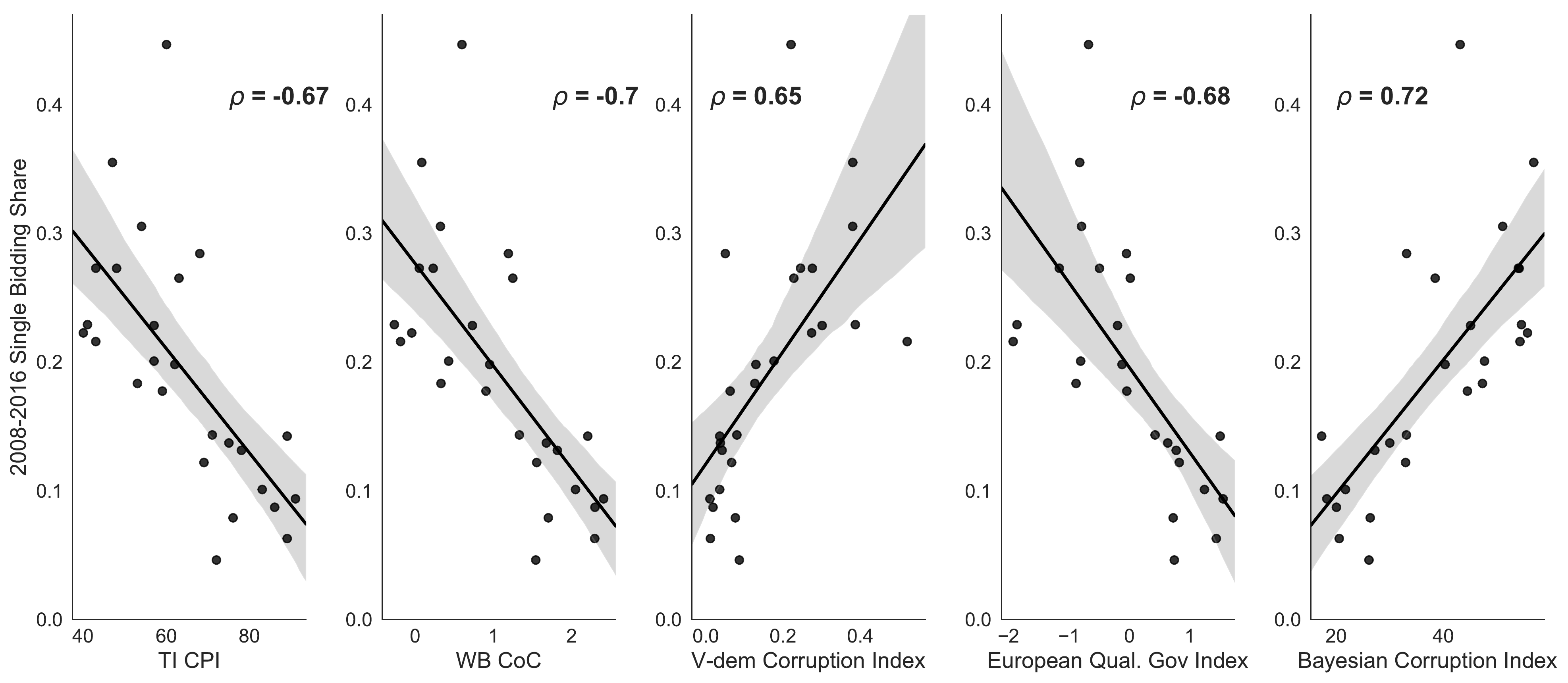}
\caption{The 2008-2016 single bidding rate of each country correlated with various perception-based measures of corruption, all measured in 2013. When correlations are negative, the perception-based measure refers to positive outcomes, for instance quality of government or control of corruption.}
\label{fig:perception_correlations}
\end{figure}

We plot the single bidding rates with of each country over the 2008-2016 period in Figure~\ref{fig:country_sb_rates}. We see significant variation, with single bidding rates below 10\% in some countries and over 30\% in others. How do these measures of corruption risk correlate with the survey-based perception indicators mentioned in the previous section? In Figure~\ref{fig:perception_correlations}, we correlate the average single bidding rates from 2008-2016 with Transparency International's Corruption Perception Index (TI CPI)~\cite{TI_CPI}, the World Bank's Control of Corruption Index (WB CoC)~\cite{WB_WGI}, Varieties of Democracy's Corruption Index (V-Dem Corruption Index)~\cite{coppedge2018v}, the Quality of Government (QoG) Institute's European Quality of Governance Index (European Qual. Gov. Index)~\cite{charron2014regional}, and the Bayesian Corruption Index (BCI)~\cite{standaert2015divining}, each measured in 2013 (this was the most recent year for which all five indicators were available). In each case we find a significant correlation between country-level single bidding rates and worse corruption or quality of government outcomes (Pearson correlation between .65 and .72).  

Our data offers several advantages over the survey-based indicators. Surveys encode subjective perceptions about corruption that may be biased. For instance, past research by Olken found that perceived corruption exceeds actual corruption when ethnic diversity is high~\cite{olken2009corruption}. This is a significant obstacle when we would like to compare the rate of corruption across countries. Moreover, the way such indicators are calculated is often modified from year to year, again making comparison difficult. From our perspective, however, the most significant advantage of the single-bidder measure of corruption risk is that it is observed at a granular level; surveys are expensive and, with few exceptions~\cite{charron2014regional}, are not carried at the sub-national, e.g. regional level. We exploit that our indicator is available at the transaction level in the following section by introducing our network science-based analytic framework.

\section{Network Analysis}
Public procurement markets can be described as bipartite networks because of the clear division of their participants into issuers and winners. Bipartite networks have been applied to the study of systems including two kinds of actors including flowers and their pollinators~\cite{jordano2003invariant}, cities and industries present in them~\cite{bustos2012dynamics}, and buyers and sellers in markets~\cite{hernandez2018trust}. We apply the same paradigm to our data on procurement markets: for each country and year in our dataset, we create bipartite networks in which issuers and winners are connected by a weighted edge, with weight counting the number of contracts between the two entities.

The standard analysis of a complex network is based on null models, which are randomized versions of the empirical system. The statistical observations made on the empirical network is compared to those obtained on the randomized versions. Significant differences can be interpreted as results of correlations, which are eliminated by the randomization process.

In Table~\ref{tab:network_summary} we report summary of the raw statistics of each country's procurement market network, averaged over 2008-2016. In all cases the average weighted degrees of nodes (counting the number of contracts the node is involved in) are smaller than their standard deviations, indicating heterogeneous degree distributions.

\begin{landscape}
\begin{table*}[t]
\begin{tabular}{lrrrrrrrrr}
\hline
Country &  \# Contracts &  \# Winners &  \# Issuers &  Density &  R-A Clust. & $ \mu(Deg_{W})$ &  $\sigma(Deg_{W})$ &  $\mu(Deg_{I})$ &  $\sigma(Deg_{I}) $\\
\hline
AT      &        3314 &      1882 &       395 &   0.0033 &    0.03 &             1.8 &             2.3 &             8.4 &            22.7 \\
BE      &        6674 &      3046 &      1039 &   0.0014 &    0.02 &             2.2 &             4.4 &             6.4 &            15.2 \\
BG      &        8653 &      2150 &       484 &   0.0048 &    0.27 &             4.0 &            14.6 &            17.9 &            56.3 \\
CY      &         916 &       403 &        64 &   0.0179 &    0.14 &             2.3 &             3.7 &            14.3 &            48.7 \\
CZ      &        8030 &      2933 &       986 &   0.0017 &    0.04 &             2.7 &             7.4 &             8.1 &            32.7 \\
DE      &       32339 &     15395 &      4049 &   0.0004 &    0.03 &             2.1 &             5.2 &             7.9 &            23.5 \\
DK      &        4858 &      2099 &       539 &   0.0028 &    0.04 &             2.3 &             4.1 &             8.9 &            26.5 \\
EE      &        1913 &       967 &       170 &   0.0083 &    0.08 &             2.0 &             2.6 &            11.0 &            28.9 \\
ES      &       20035 &      7496 &      1765 &   0.0011 &    0.13 &             2.7 &             6.8 &            11.3 &            31.8 \\
FI      &        6248 &      2750 &       578 &   0.0029 &    0.05 &             2.3 &            12.2 &            10.8 &            27.4 \\
FR      &      120946 &     42562 &      6294 &   0.0003 &    0.08 &             2.8 &            11.5 &            19.3 &            51.4 \\
GR      &        4246 &      2348 &       437 &   0.0031 &    0.12 &             1.8 &             2.2 &             9.7 &            44.0 \\
HU      &        5700 &      2016 &       610 &   0.0026 &    0.08 &             2.8 &             5.9 &             9.4 &            27.4 \\
IE      &        2713 &      1587 &       208 &   0.0056 &    0.03 &             1.7 &             2.3 &            13.1 &            51.3 \\
IT      &       18249 &      7749 &      2434 &   0.0006 &    0.07 &             2.4 &             7.2 &             7.5 &            29.0 \\
LT      &        9007 &      1368 &       272 &   0.0084 &    0.32 &             6.5 &            40.2 &            32.5 &           178.2 \\
LV      &        9451 &      2148 &       262 &   0.0057 &    0.15 &             4.3 &            14.8 &            36.1 &           119.4 \\
NL      &        6691 &      3579 &      1136 &   0.0013 &    0.02 &             1.8 &             2.6 &             6.0 &            14.7 \\
NO      &        3479 &      1899 &       497 &   0.0031 &    0.04 &             1.8 &             2.7 &             7.0 &            12.6 \\
PL      &      108886 &     19079 &      3649 &   0.0006 &    0.23 &             5.7 &            64.7 &            29.7 &            96.9 \\
PT      &        2255 &      1052 &       334 &   0.0041 &    0.08 &             2.1 &             3.1 &             6.7 &            30.4 \\
RO      &       19807 &      3503 &       939 &   0.0025 &    0.32 &             6.0 &            44.5 &            21.1 &            72.5 \\
SE      &        9441 &      4721 &       724 &   0.0022 &    0.06 &             2.0 &             3.7 &            13.1 &            33.3 \\
SI      &        6623 &      1268 &       448 &   0.0067 &    0.27 &             5.2 &            18.7 &            14.8 &            33.3 \\
SK      &        2654 &      1068 &       344 &   0.0043 &    0.09 &             2.4 &             5.1 &             8.1 &            32.5 \\
UK      &       32275 &     15577 &      2230 &   0.0006 &    0.04 &             2.1 &             3.7 &            14.6 &            50.0 \\
\hline
\end{tabular}
\caption[Procurement Market Network Summary Statistics]{Summary statistics of procurement market networks, averaged over 2008-2016. R-A Clust. refers to Robins-Alexander clustering ~\cite{robins2004small}, a measure of the local correlation of connectivity in bipartite networks, analogous to the clustering coefficient in monopartite networks. The final four columns present the averages and standard deviations of winner and issuer strengths, their degrees weighted by contract count, respectively.} 
\label{tab:network_summary}
\end{table*}
\end{landscape}

\begin{figure*}[!t]
\centering
  \includegraphics[width=\textwidth]{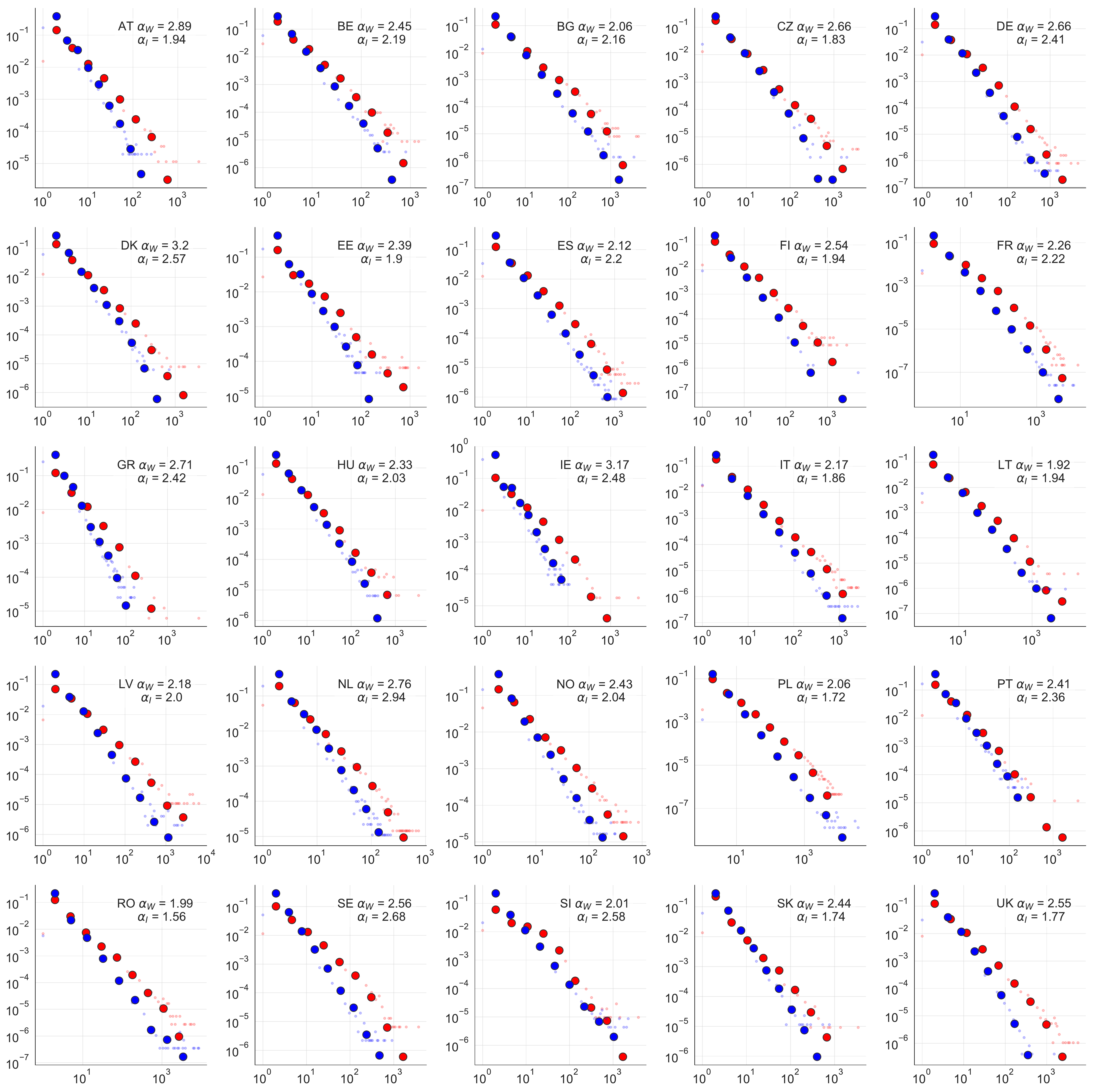}
  \caption[Contract distributions of issuers and winners, Country-level.]{The distribution of the number of contracts awarded and won, for issuers (in red) and winners (in blue), respectively, of procurement contracts in each country, aggregated from 2008 to 2016. The distributions are plotted on a log-log scale. We report the alpha parameter of a power-law degree distribution fitted to both distributions in the plot~\cite{alstott2014powerlaw}.}
  \label{fig:national_dists}
\end{figure*}

We inspect the degree heterogeneity of each country graphically in Figure~\ref{fig:national_dists}. Plotted on a log-log scale the degree distributions of both issuers and winners are heterogeneous in all countries: some rare issuers and winners are involved in hundreds, even thousands of contracts across the time span of our data, while the majority are involved in only a few contracts. This heterogeneity cannot be considered as a signature of any corruption, it is more related, e.g., to the broad size distribution of firms and institutions~\cite{Axtell2006frimsize}.

We also report the Robins-Alexander clustering of each network in Table ~\ref{tab:network_summary}. Robins-Alexander clustering is defined as the number of cycles of length four in the network divided by the number of paths of length three~\cite{robins2004small}. Given that a firm wins contracts from two issuers, and that another firm wins a contract from one of these two issuers, Robins-Alexander clustering can be interpreted as the probability that this second firm also wins a contract from the other issuer. This measures the tendency for local clustering in the market analogous to the clustering coefficient in monopartite networks. The expected Robins-Alexander clustering of random bipartite networks tends to their density as they get large~\cite{robins2004small}. As seen in Table~\ref{tab:network_summary}, the observed clustering is typically an order of magnitude greater than the density of the observed networks. This indicates the presence of significant local correlations in the markets.  

These descriptive statistics indicate that our networks have rich structure which deviates significantly from random behavior. In the next subsections we introduce two measures of the structure of the market: its centralization and the extent too which it is clustered. We then exploit these measures to describe the distribution of corruption risk in each market.

\subsection{Core-Periphery Analysis}
A common task of network analysis is to highlight the most central and active nodes. Some empirical networks have a group of densely connected actors at their centers, sometimes called \textit{cores}~\cite{csermely2013structure}. Besides highlighting important actors, methods to detect network cores can be used to compare the degree of centralization in a network. In this subsection, we adapt a method known as k-shell decomposition to weighted bipartite networks in order to rank issuers and winners by their centrality in the network. We say that the most central actors form the core of the market. We measure and compare the relative sizes of the cores of each procurement market, interpreting a larger core as a signal of procurement market centralization. We find a strong correlation between market centralization and corruption risk. 

In generic unweighted graphs, the concept of a coreness is defined iteratively~\cite{batagelj2003m}. To start, all nodes of degree 1 are assigned a core number of 1 and then removed from the graph. In the next step, all remaining nodes with degree 1 in the trimmed graph are also assigned core number 1 and then removed. This process repeats until all nodes have at least degree 2. Nodes with degree 2 are assigned core number 2 and they are the first nodes to be removed in the next iteration of removals. This process yields a hierarchical decomposition of the nodes in the network by their core number. The k-core of the graph refers to the subgraph consisting of nodes with core number at least k~\cite{dorogovtsev2006k}. 

Our networks have two features that distinguish them from the graphs considered in this example. The first is that the edges include weights, encoding the volume of contracts between the focal issuer and winner. The second aspect is that the different node sets have different degree distributions. We tweak the concept of the core number and k-core in order to apply it to our networks.

Our method modifies the iterative procedure described above to consider the weighted degrees, defined as the count of contracts they are involved in as either issuers or winners, respectively, of nodes. Otherwise the method is identical. As the edge weights in our networks are integers (describing contracting volume), no further modification is necessary. This is a simplified version of the weighted k-shell decomposition of Garas et al.~\cite{garas2012k}.

\begin{figure*}
\centering
  \includegraphics[width=\textwidth]{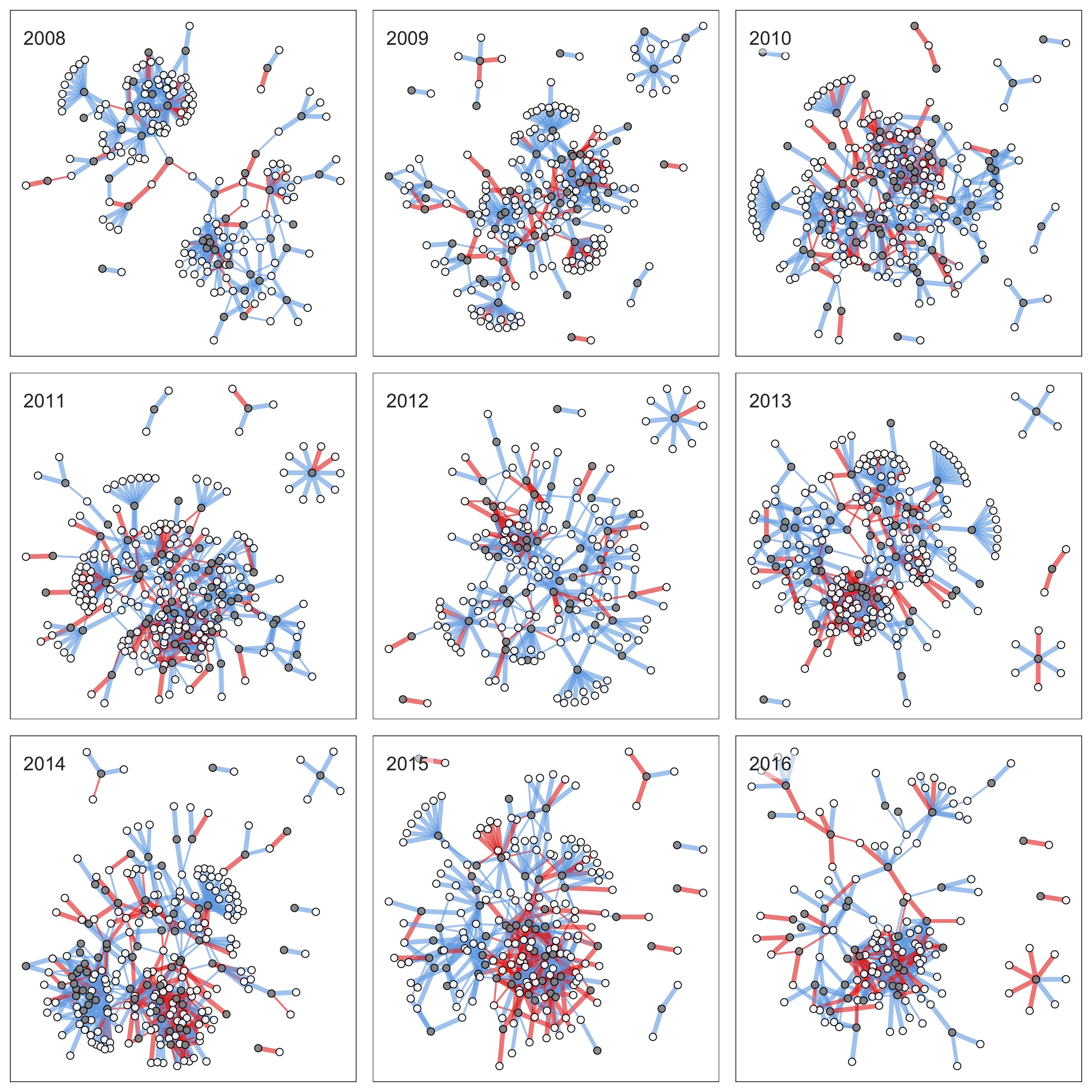}
  \caption[Hungarian Market Cores]{The weighted core of the Hungarian market from 2008 to 2016. Gray nodes denote issuers of contracts, while white nodes denote winners. Nodes are included in the core if they engage in many contracts with other highly active nodes, defined iteratively. Edges are colored red if the rate of single bidding on contracts between the issuer and winner exceeds the average single bidding rate observed in the whole market (including the periphery). Note that the core index of each node considers edge weights encoding the count of contracts between issuers and winners.}
  \label{fig:hu_core}
\end{figure*}

This procedure assigns each node in our network a weighted core number, measuring its centrality in the network. As we would like to partition the network into a core and periphery, we need to determine a cutoff for a node to be considered a member of the core. This cutoff should be a function of the size of the network as it would not be appropriate to use the same cutoff for networks of vastly different sizes. We also address the concern that issuers and winners have different degree distributions by setting two different cutoffs for core membership: one for issuers and one for winners. In the end we choose to consider issuers (respectively winners) as core issuers if their weighted core number exceeds the average degree of issuers (respectively winners) of the network.

We visualize the Hungarian core in Figure~\ref{fig:hu_core}, marking edges with above market average single-bidding rates in red. The visualization highlights the importance of considering the volume of contracting between issuers and winners in our measure. Issuers and winners can have only a single neighbor and still be considered for core membership because they may have a high volume of contracts with that neighbor.

We observe that the core has a consistent size over time, and remains densely connected in all years. We confirm that in general the weighted core subgraph has a much higher density in a summary statistics table included in the appendix. In general between 20 to 60\% of all contracts awarded in a given country are between core issuers and winners.

\begin{figure*}
\centering
  \includegraphics[width=\textwidth]{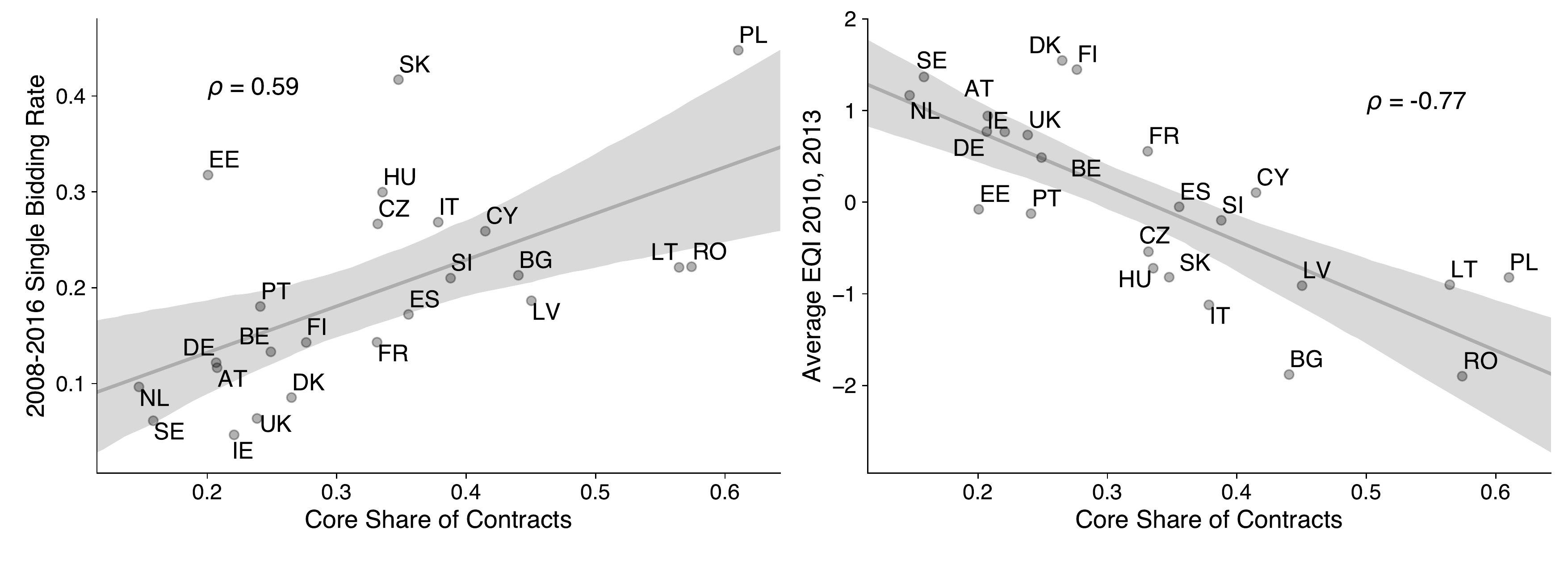}
  \caption[Market Centralization and Corruption]{Comparing the centralization of the procurement markets, measured as the share of all contracts which are between core issuers and winners, and corruption risk quantified by single-bidding and the European Quality of Government Index (EQI), respectively. All measures are averaged over the years 2008-2016. We report Pearson correlations, both significant at $p<.01$ and errors indicate 95th percentile bootstrapped confidence intervals.}
  \label{fig:centralization_vs_sb_eqi}
\end{figure*}

We now pose two questions about the relationship between network centralization and corruption risk measured by single-bidding. First: what is the relationship between the degree of centralization of a market and its corruption risk. In Figure~\ref{fig:centralization_vs_sb_eqi} we plot the relationship between centralization, quantified by the share of all contracts between core issuers and winners, against corruption, measured by the overall single bidding rate and the survey-based EQI measure. We find a significant positive relationship between procurement market centralization and corruption. In the political economy literature there is an ongoing debate about the relationship between the centralization of government power and corruption. Our finding supports the side of the debate claiming that centralization facilitates corruption~\cite{persson2002political}.

\begin{figure*}
\centering
  \includegraphics[width=\textwidth]{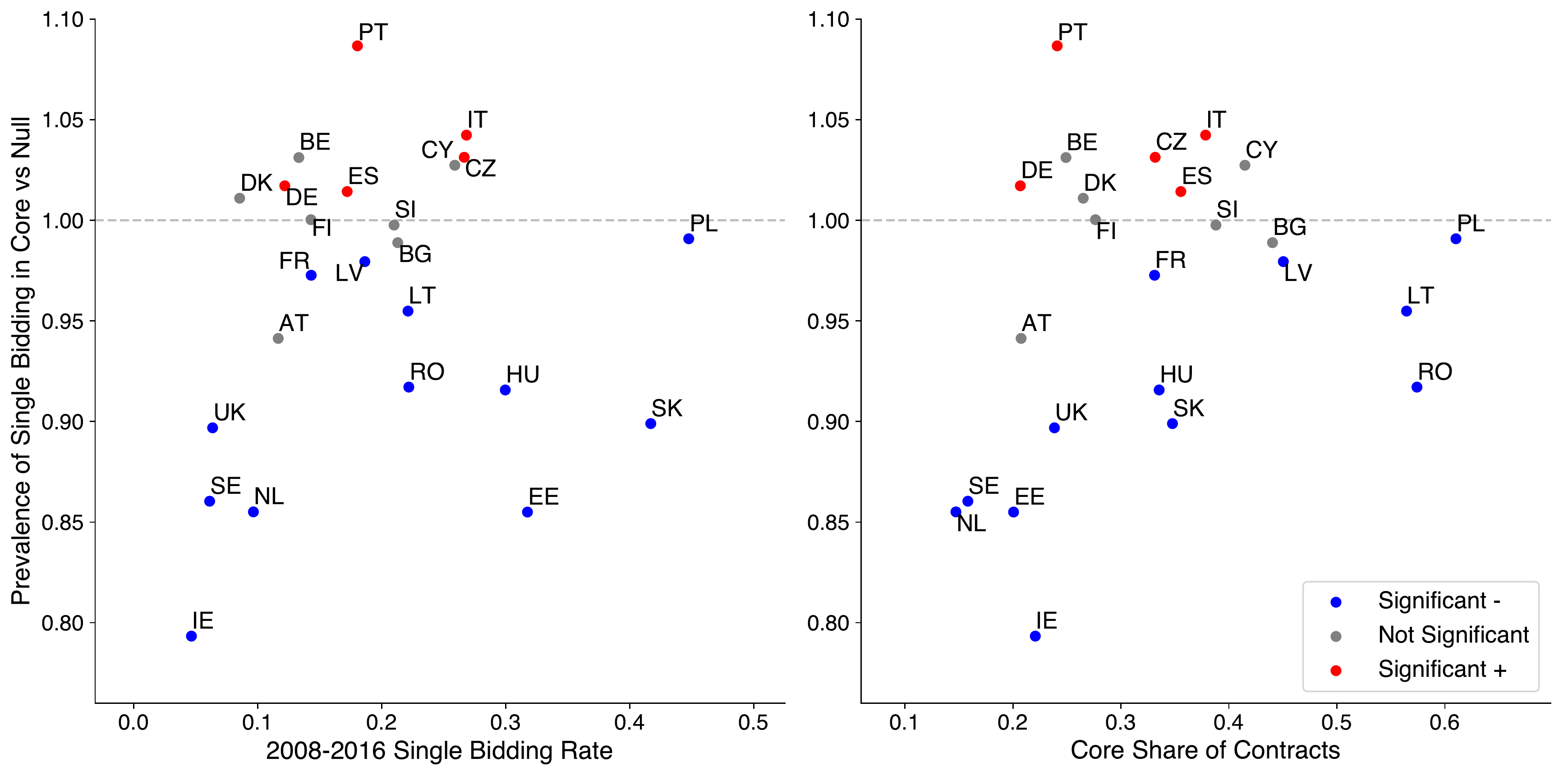}
  \caption[Prevalence of single bidding in cores.]{Comparing the relative prevalence of single bidding in the core of EU procurement markets with their overall single bidding rates and the relative core sizes, respectively. Blue points are countries in which single bidding is less common among core contracts than expected under a sector-preserving null model. Red points are countries in which the opposite is true: single bidding is significantly more common in core contracts.}
  \label{fig:core_sb}
\end{figure*}

We are also interested in the distributional nature of corruption risk. Does centralization induce corruption by fostering corruption among core issuers and winners? We probe this question by comparing the rate of single-bidding in the core against a null model that randomizes the distribution of single-bidding. Specifically, we preserve network structure and the label of nodes as core or periphery, but randomly shuffle the single-bidding labels. We restrict the randomization by permuting single-bidder labels only across contracts with the same two-digit CPV code. This takes into account market specific effects. For each market we divide the observed rate of single-bidding in its core $SB_{core}$ by the average rate of single-bidding in its core over 1000 CPV-preserving randomizations $\mu(SB_{core}^{rand})$. Statistical significance of any deviation between observation and the distribution resulting from the randomizations are determined by calculating a z-score. We plot the results averaged over all years for each country in Figure~\ref{fig:core_sb}.

Unlike in the previous figure which indicated a strong relationship between market centralization and corruption risk, no clear pattern emerges. In some countries corruption risk in the form of single-bidding is significantly over-represented in the core. In other countries it is significantly underrepresented. For example, Czech Republic and Hungary have similar overall corruption risk scores, but in the former single-bidding is significantly more common between core issuers and winners, while in the latter it is more common between actors on the periphery. In the second panel of Figure~\ref{fig:core_sb} we show that the size of the core does not have any significant relationship with the over-representation of corruption risk in the core.

We draw several conclusions from this analysis. The first is that, as mentioned, there is a strong correlation between centralization in procurement markets and their overall procurement risk. What is less clear is by what mechanism this effect might emerge. Our latter findings show that centralization is not related to higher corruption risk in either the core or periphery \textit{in general}. The observed heterogeneity in the tendency of corruption risk to accumulate in the core or periphery of different countries has important policy implications because the strategies used by corrupt actors to extract rents are likely very different.

\subsection{Edge Clustering}

\begin{figure*}
\centering
  \includegraphics[width=\textwidth]{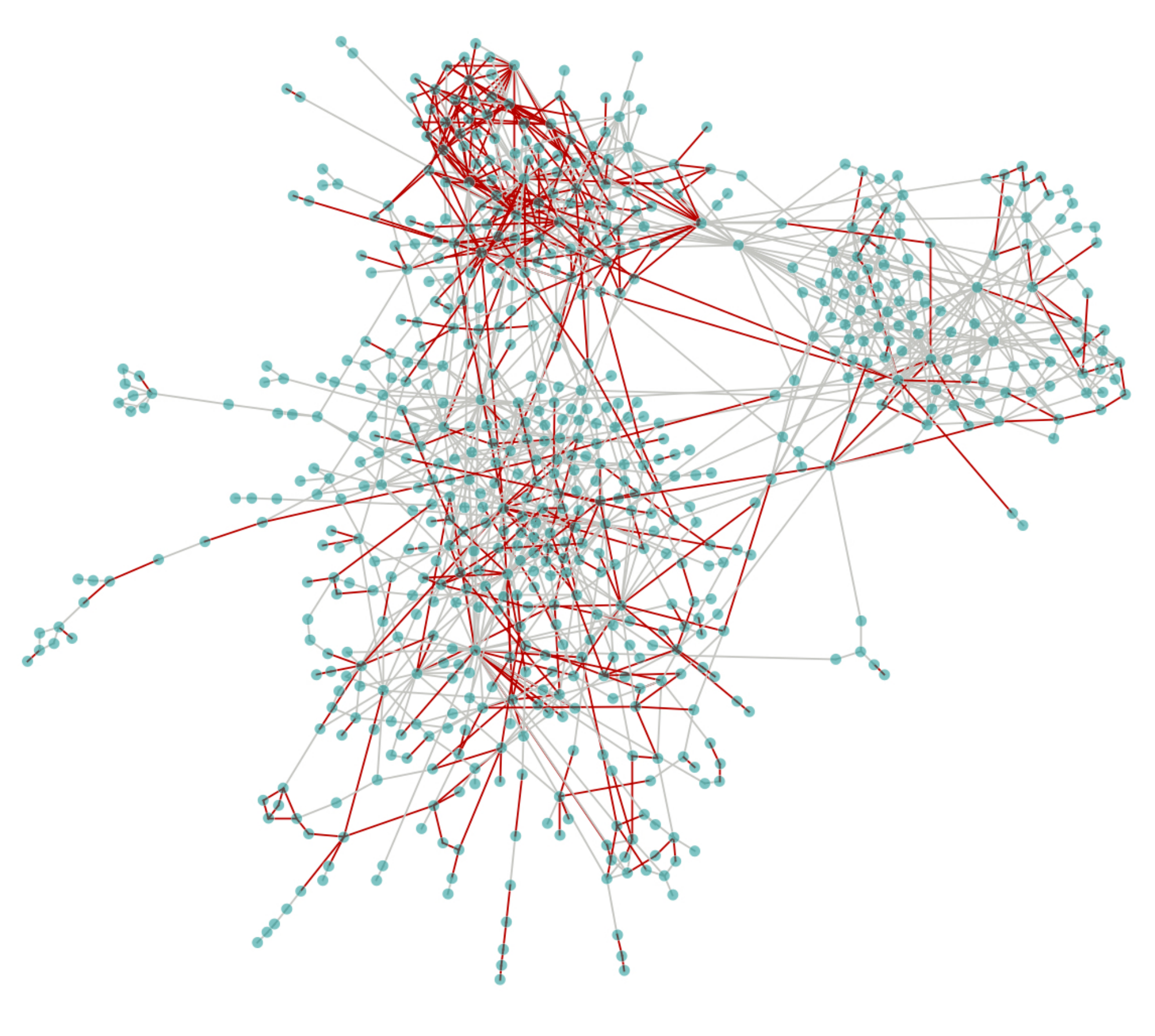}
  \caption[2014 Hungarian Procurement Market with Edge Communities]{The 2014 Hungarian procurement market. We plot the largest connected component of the network, filtering out nodes involved in less than three contracts for the sake of visualization. Nodes are buyers and suppliers of contracts, connected by an edge if they contract with one another. Edges are colored red if the single bidding rate on the edge exceeds the average rate of single-bidding that year. Single bidding is significantly over-represented among the edges in the top left cluster.}
  \label{fig:hu2014_edgecluster}
\end{figure*}

Another dimension of potential heterogeneity in the distribution of corruption risk is its clustering. Previous research has found that corruption is significantly correlated in the network, with high corruption risk edges typically bunched together~\cite{fazekas2016corruption}. To demonstrate this idea, we plot the Hungarian procurement market in 2014 in Figure~\ref{fig:hu2014_edgecluster}. Coloring edges with above-market average rates of single bidding, a clear pattern emerges: the top left cluster of nodes has significantly higher rates of single bidding than other parts of the network. We propose to quantify this tendency, as well as the overall tendency of a procurement market to exhibit topological clustering using community detection. We first group the edges of the network into communities and measure the quality of this partition. We then calculation the variation of single bidding rates across communities as a measure.

Community detection is a distinguished area of research interest in network science~\cite{fortunato2010community}. There are many methods to partition the nodes of networks into communities. As our object of interest, corruption risk, is an attribute of network edges, we should rather partition the edges of the network into communities. There are several well-studied methods used to cluster the edges of networks into so-called ``link-communities''. We adopt one approach based on line graphs~\cite{evans2010line,ahn2010link}. The line graph of a network is the network obtained if the edges if the original graph are considered as nodes and then connected if they share a node in the original graph. We transform each of our networks into the corresponding line graph, obtaining a new network for each one in which the nodes correspond to the original network's edges or contract relationships. We can then apply any standard clustering algorithm on the new network to obtain an assignment of each issuer-winner edge to a community. We apply the Louvain method, a computationally fast and accurate method commonly used to detect communities in networks~\cite{blondel2008fast}.

We can also use the partition of the line graphs into communities to describe the extent to which the networks are topologically clustered. We measure the tendency of edges to be within rather than between communities detected by the Louvain method using \textit{modularity}, a quality function of network partitions. Modularity varies between -1 and 1, with negative scores indicating that edges are rather present between "communities" of the given partition than within them, scores around 0 indicating that there is no difference in the frequency of inter vs intra-community edges, and higher values indicating that edges are much more likely to be between nodes of the same community rather than across communities. In other words, higher modularity scores indicate that the network has more distinct and separated groups of nodes.

\begin{figure}[t]
  \includegraphics[width=\textwidth]{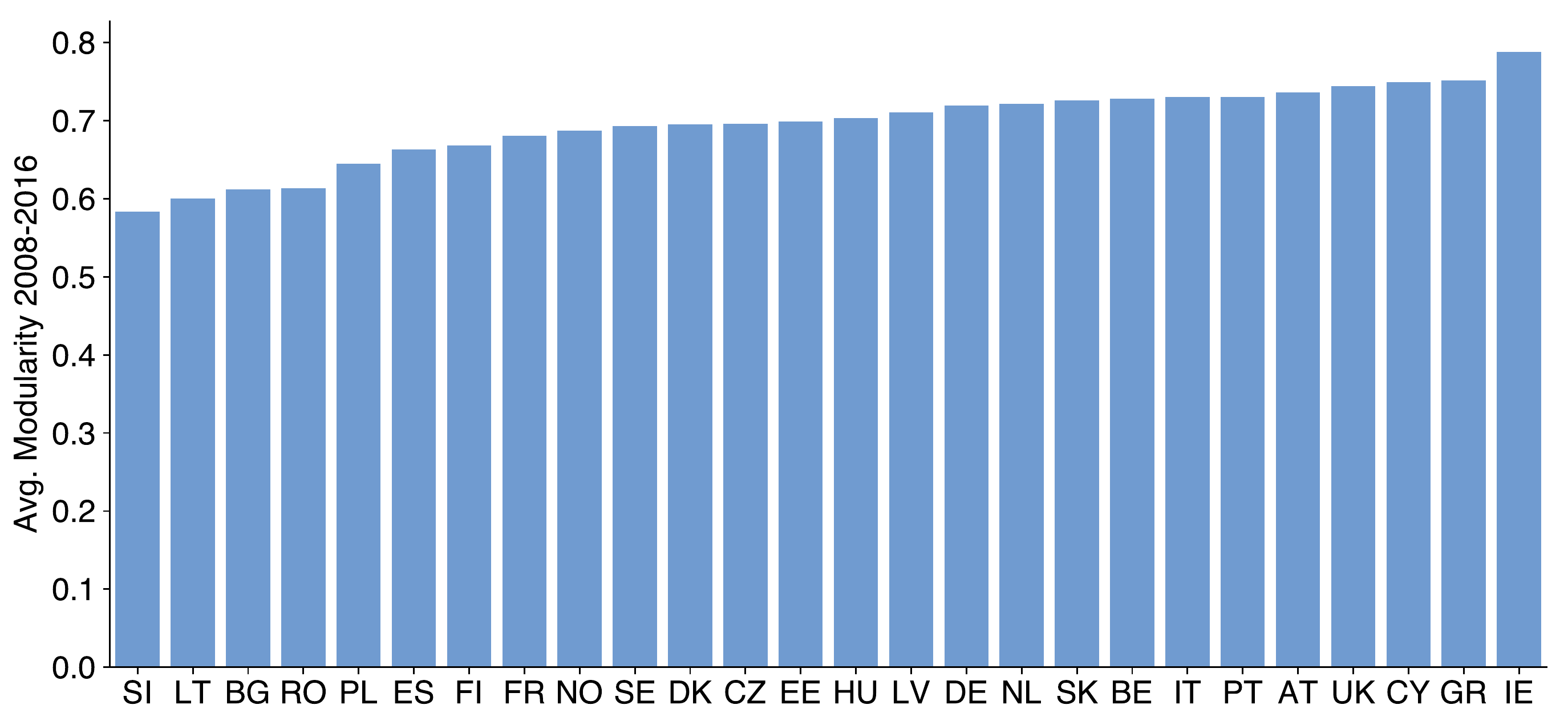}
\caption{Modularity scores of procurement market networks by country, averaged over 2008-2016. Higher values indicate more significant topological clustering in the market.}
\label{fig:country_modularities}
\end{figure}

We plot each country's average modularity score from 2008 to 2016 in Figure~\ref{fig:country_modularities}. There is significant topological clustering in all the networks in our data, similar to what can be observed in our plot of the Hungarian market in 2014 (modularity ~.71). Given the partition of the edges of each network into communities, we can now quantify the extent to which single bidding varies across the different communities of a network. 

The partition of edges in a network, denoting contracting relationships between issuers and winners naturally gives us a partition of all contracts awarded in a given market. We then calculate the coefficient of variation of single-bidding across the clusters, defined as the standard deviation of the single-bidding rates across the contract clusters over the average. As clusters can have significantly different numbers, we actually calculate weighted standard deviation $\sigma^{W}_{SB}$ and mean $\mu^{W}_{SB}$ of single bidding across clusters, defined as follows:

$$\sigma^{W}_{SB} = \sqrt{ \frac{ \sum_{c\in C} |c| (sb_{c} - \mu^{W}_{SB})^2 }{ \frac{(|C|-1)}{|C|} \sum_{c\in C} |c| } },$$
and
$$\mu^{W}_{SB} =  \frac{\sum_{c \in C} |c| sb_{c}}{\sum_{c \ in C} |c|},$$

\noindent where $C$ denotes the set of contract clusters, $c$ is a specific cluster, and $sb_{c}$ is the rate of single bidding in the cluster $c$. The weighted coefficient of variation, which in our context we refer to as the clustering of single bidding, is simply the ratio $\sigma^{W}_{SB}/\mu^{W}_{SB}$.

\begin{figure}[t]
  \includegraphics[width=\textwidth]{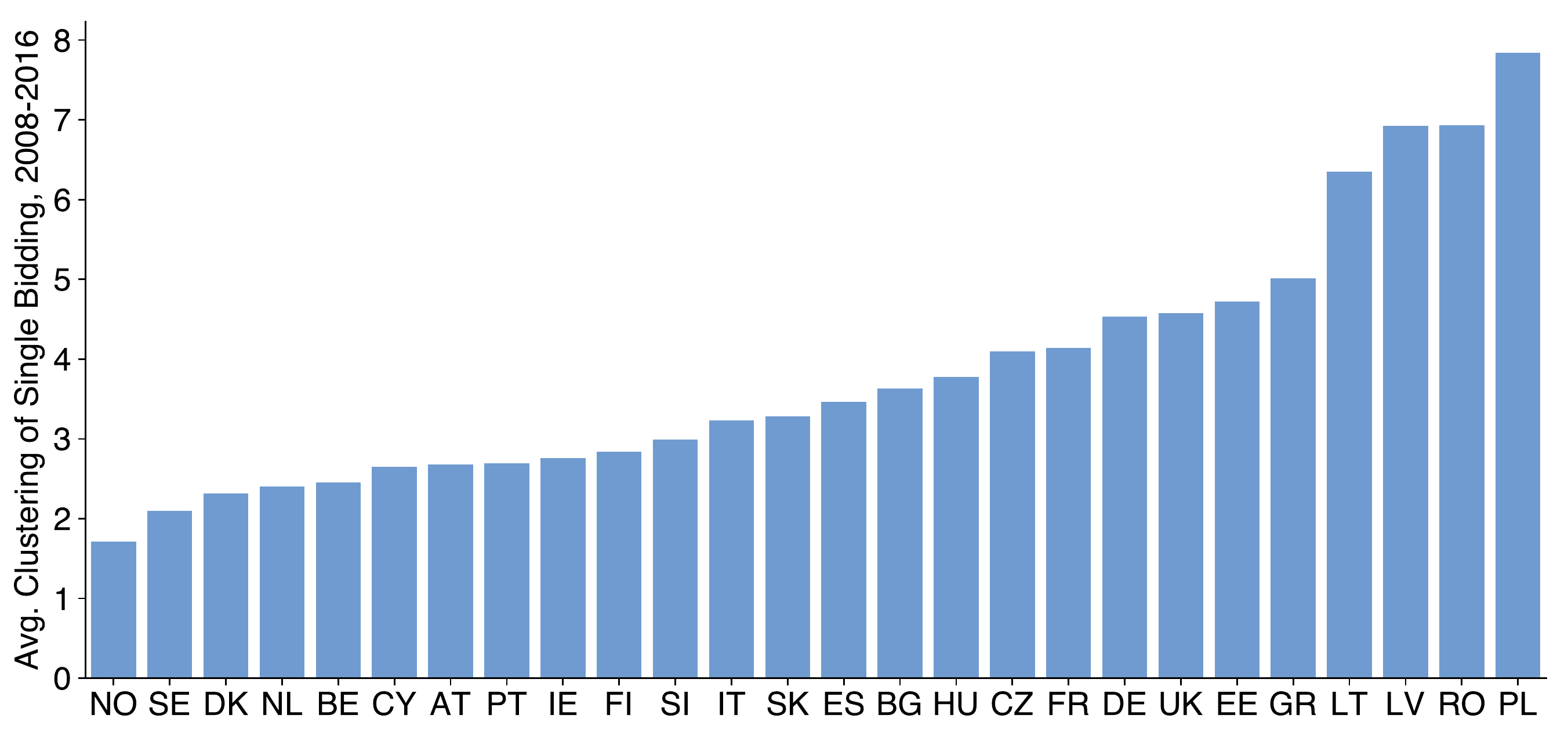}
\caption{Clustering scores of procurement market networks by country, averaged over 2008-2016. Higher values indicate higher variation of single-bidding rates across communities in the market compared to a shuffled null model.}
\label{fig:country_sb_clustering}
\end{figure}

As with our measure of centralization, we compare our observed clustering of single bidding measure against a suitable null model. We again opt to randomize the single-bidder label on contracts within the 2-digit CPV classes to create a realistic yet randomized distribution to compare against the empirical data. 1000 times we shuffle the single bidder label on all contracts and recalculate the clustering of single bidding for these randomized markets. We divide the observed clustering of single bidding by the average of the 1000 randomized clustering of single bidding scores. We plot the result by country, averaged over 2008-2016, in Figure~\ref{fig:country_sb_clustering}. In every case single bidding is non-trivially clustered within communities in the network. The magnitude of observed clustering ranges from 2 to over 8 times higher than expected in the sector-preserving randomization. We interpret this as significant evidence that corruption is not randomly distributed in the public sector. 


The extent to which corruption risk clusters in the market has important policy implications. The greater the clustering of single bidding in a market, the more likely it is that investigations of the network neighbors of known corrupt actors will be successful. For example, returning to our visualization of the Hungarian market in 2014 (see Figure~\ref{fig:hu2014_edgecluster}), we do not only suggest that corruption risk is significantly higher in the northwestern community, but also that investigators of corruption in Hungary should consider this pattern of clustering in the future to shape their strategy.

\begin{figure}[t]
  \includegraphics[width=0.9\textwidth]{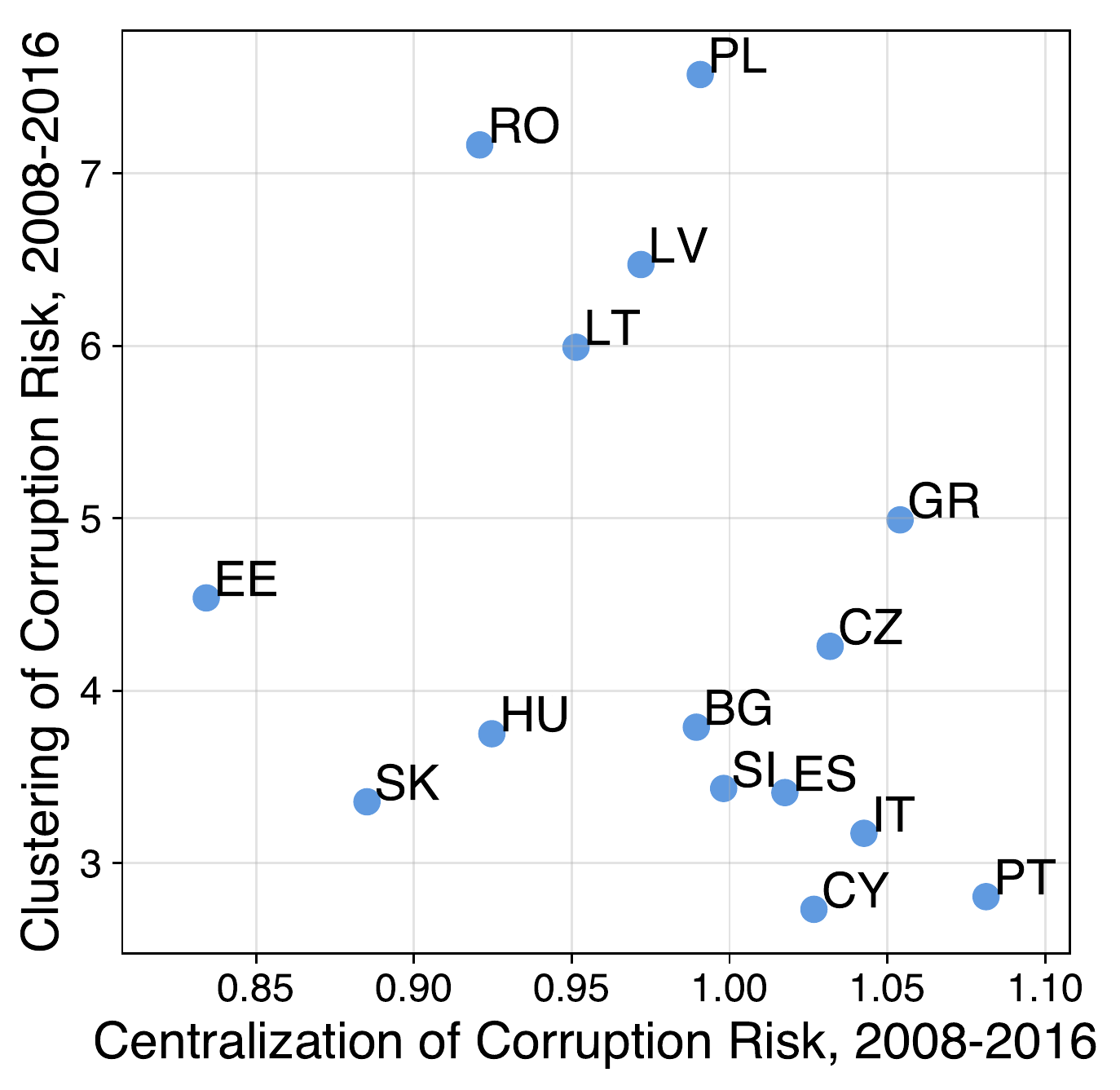}
\caption{The relative prevalence of corruption risk in the core of the market plotted against its tendency to cluster, plotted for countries with above average single bidding rates in the EU, 2008-2016.}
\label{fig:cent_vs_clust}
\end{figure}

We now compare our two measures of the distribution of corruption risk in markets. For simplicity we consider only those countries with above average single bidding rates in our data. We plot the centralization of corruption risk against its tendency to cluster for these countries in Figure~\ref{fig:cent_vs_clust}. We observe that although corruption risk is high in all countries, corruption risk within each country is distributed differently. Countries in the bottom right corner of the plot, such as Portugal and Italy, have higher corruption risk in the core of their markets and a weaker tendency for risk to cluster. Countries in the top and center of the plot such as Poland and Latvia have a neutral distribution of risk across the core and periphery of their markets, but overall risk has a strong tendency to cluster. Finally, countries such as Hungary, Slovakia, and Estonia have higher corruption risk in their peripheries and a moderate tendency for risk to cluster.

This emphasis on the distribution of corruption risk presents a novel way to think about corruption in a comparative manner. Though it may not be accurate to say that each country with high corruption risk is corrupt in its own way, these deviations suggest that corruption may be organized in different ways, likely reflecting political and economic constraints. The differences in topological structure (i.e. the degree of centralization, measured by the share of contracts in the core of the market, and the clustering of edges into communities measured by modularity) of the markets also indicate that interesting structures emerge in the mesoscopic level of this data: between the micro-level of individual transactions and the global picture of the entire market.

\section{Discussion and Future Work}

In this paper we applied network science methods to mine a big administrative dataset on public procurement contracts for insights on the distribution of corruption risk. We found that countries may have similar levels of corruption risk but significantly different distributions of that risk. In some countries corruption risk is more prevalent in the center of the network, while in others among peripheral actors. We also found that the degree of centralization in procurement markets overall is a strong predictor of their global corruption risk and low quality of government. Another dimension of heterogeneity is the degree of clustering of corruption risk: though corruption risk is significantly clustered in all countries in our data, the extent of the clustering varies between 2 to 8 times what is expected under a sector-preserving randomization.

These heterogeneities are not merely curious artifacts in the data. They have significant implications for anti-corruption policy. For example, when corruption risk is centralized, it is likely that the central government itself is a hotbed of corruption, and should not be trusted to address the problem. If corruption risk is highly clustered, successful investigations of corruption should snowball by following up with ``nearby'' actors. 

We suggest several directions to extend this work. For instance, considering that corruption is known to vary considerably at the regional level~\cite{charron2014regional}, geographic information about the locations of issuers and winners could enhance our analysis~\cite{popa2019uncovering}. Our approach aggregates both corruption risk and network structure over time - certainly it is the case that corruption can change over time. Past work on the response of corrupt networks to political turnover suggests that procurement markets are significantly rewired following changes of government~\cite{fazekas2017turnover}. Comparative analyses and case studies using this data and approach promise to enhance our understanding of the different ways corruption works, and potentially how it can be limited.

\section*{Acknowledgements}
The authors would like to thank \'Agnes Batory and Eelke Heemskerk for useful feedback on a preliminary version of this paper. JK acknowledges support from the Hungarian Scientific Research Fund (OTKA K129124 - ”Uncovering patterns of social inequalities and imbalances in large-scale networks”).

\section*{Conflict of Interest}

The authors declare no conflict of interest.

\bibliographystyle{spphys}       
\bibliography{ds_corr.bib}   

\newpage

\section{Appendix}
Here we report additional information about the cores of the procurement markets in each country.
\begin{table}[!htbp]
\begin{tabular}{lrrrrrr}
\hline
Country &  Core \#Contracts &  Share &  \#Winners &  \#Issuers &  \#Edges &  \%Single bid.  \\
\hline
AT      &              696 &                 0.21 &            111 &             25 &          168 &          0.14 \\
BE      &             1680 &                 0.25 &            188 &             92 &          382 &          0.14 \\
BG      &             3923 &                 0.44 &            162 &             61 &         1287 &          0.18 \\
CY      &              390 &                 0.42 &             32 &              3 &           41 &          0.39 \\
CZ      &             2764 &                 0.34 &            218 &             88 &          663 &          0.26 \\
DE      &             6973 &                 0.21 &            802 &            229 &         1827 &          0.16 \\
DK      &             1342 &                 0.27 &            147 &             45 &          323 &          0.11 \\
EE      &              441 &                 0.21 &             63 &             12 &          120 &          0.19 \\
ES      &             7182 &                 0.36 &            529 &            158 &         3206 &          0.18 \\
FI      &             1712 &                 0.27 &            167 &             50 &          661 &          0.15 \\
FR      &            39462 &                 0.32 &           2829 &            582 &        16345 &          0.14 \\
GR      &              715 &                 0.18 &            122 &             23 &          267 &          0.26 \\
HU      &             1949 &                 0.34 &            163 &             56 &          412 &          0.27 \\
IE      &              682 &                 0.24 &             91 &              9 &          111 &          0.02 \\
IT      &             6958 &                 0.38 &            462 &            156 &         1980 &          0.29 \\
LT      &             5414 &                 0.57 &             94 &             27 &          711 &          0.17 \\
LV      &             4664 &                 0.46 &            161 &             27 &          481 &          0.18 \\
NL      &             1137 &                 0.16 &            176 &             63 &          344 &          0.08 \\
NO      &              552 &                 0.15 &             87 &             28 &          233 &          0.06 \\
PL      &            66963 &                 0.61 &            896 &            455 &        14483 &          0.44 \\
PT      &              653 &                 0.26 &             67 &             20 &          116 &          0.22 \\
RO      &            12087 &                 0.59 &            165 &            106 &         2367 &          0.14 \\
SE      &             1761 &                 0.17 &            268 &             33 &          686 &          0.04 \\
SI      &             2632 &                 0.39 &             90 &             84 &          976 &          0.19 \\
SK      &              934 &                 0.34 &             68 &             28 &          144 &          0.37 \\
UK      &             8511 &                 0.25 &           1072 &            119 &         2058 &          0.08 \\
\hline
\end{tabular}
\caption[Summary statistics of market cores.]{The core statistics of each national market, averaged over 2008-2016. Core share refers to the share of overall contracts awarded that are between core issuers and winners.}
\label{tab:core_stats}
\end{table}

\end{document}